\documentclass[preprint]{aastex}

\shorttitle{Giant Pulses from the Crab Pulsar}
\shortauthors{Alexander ~A. ~Ershov}

\begin{document}

\title{
On scattering of Giant Pulses from the Crab Pulsar: a Scattering
Function
}

\author{Alexander A. Ershov \altaffilmark{1,2}}

\altaffiltext{1}{Pushchino Radio Astronomy Observatory, Astro Space
Center, Lebedev Physical Institute, Russia}

\altaffiltext{2}{E-mail:  ershov@prao.ru}

\begin{abstract}
Simultaneous dual-frequency observations of giant pulses from the
Crab pulsar were performed at the frequencies of 61 and 111 MHz.
It is shown that scattering of giant pulses from the Crab pulsar
occurs at thick, and not at thin screen.
\end{abstract}

%\bigskip

\section{Introduction}

The Crab pulsar was discovered through detection of its giant
pulses (\cite{sta68}). Observations of giant pulses from the Crab
pulsar provide a good possibility for investigation of
interstellar medium. Observed profile of giant pulse is the
convolution of intrinsic pulse profile, function of scattering in
interstellar medium and functions related with instrumental
broadening of the profile (dispersion broadening and time
integration). Scattering in interstellar medium (10 to 15~ms at
the frequency of 111~MHz) leaves only millisecond time scale for
investigations at low frequencies, and in this case intrinsic
profile of giant pulse from the Crab pulsar may be considered as a
delta-function. Therefore, at minimization of instrumental
broadening an observed giant pulse profile will be function of
scattering. To determine function of scattering (scattering on a
thin or a thick screen, in particular) simultaneous dual-frequency
observations of giant pulses from the Crab pulsar were performed.

\section{Observations and Data Reduction}

The observations were performed since April 1 to April 7 2007
(five observing sessions) with BSA radio telescope (at 111~MHz)
and DKR-1000 radio telescope (at 61~MHz) of Pushchino Radio
Astronomy Observatory. One linear polarization was received. I
used Pulsar Machine with Fast Fourier Transform (512 channels in
2.5~MHz band and a central frequency of 110.83~MHz) and
128-channel receiver with a channel bandwidth of 20~kHz and a
center frequency of 61.39~MHz. The sampling intervals were 1.024
and 8.192~ms; durations of one observing session were 3.4~min and
16.6~min; dispersion broadenings in one channel band for a
dispersion measure of $DM = 56.759~\rm pc~cm^{-3}$ were 1.69~ms
and 40.7~ms for 111 and 61~MHz frequencies, respectively.
Dispersion measure is as referred in Jodrell Bank Crab Pulsar
Monthly
Ephemeris~\footnote{http://www.jb.ac.uk/$\sim$pulsar/crab.html}.

Upon observations at low frequencies with linear polarized
antennae effect of rotation of radio emission polarization plane
(Faraday effect) should be considered. The rotation measure of the
Crab pulsar is $RM = -42.3~\rm rad/m^2$ and this causes periodic
modulation of a signal with periods of 563 and 95.6~kHz at the
frequencies of 111 and 61~MHz respectively. Therefore, at building
of giant pulse profiles not all frequency channels were used, but
461 channels at 111~MHz and 125 channels at 61~MHz, only; this
corresponds to 4 and 26 full periods of Faraday modulation. Thus,
though observations were performed on linearly polarized antennae,
resulting profiles of giant pulses are equivalent to those,
obtained at total power mode.

Fig.~1 presents profiles of 14 giant pulses at 111~MHz. Signal to
noise ratios of all these pulses are 20 or higher. The strongest
pulse had a signal-to-noise ratio about of 150. Exponential tail,
resulting scattering of emission in interstellar medium is
distinctly seen in all profiles.

\section{Results}

Difference between scattering on thin and thick screens is
demonstrated by two effects: duration of front edge (zero for a
thin screen) and additional delay of pulse maximum at lower
frequencies in case of a thick screen.

Fig.~2 presents fine structure of front edge for 14 giant pulses,
full profiles of which are presented at the fig.1. Mean duration
of front edge of these 14 pulses is approximately 6~ms.
Instrumental broadening at this frequency is distinctly less, and
it equals to 2.0~ms (two points on a figure). Thus, duration of
front edge of scattering function (rise time) is not zero one.

Fig.~3 presents profiles of 15 giant pulses at the frequency of
61~MHz. The scattering tails are also clearly visible.

Preliminary results of processing of the observations of giant
pulses result as follows: scatter broadening values $\tau_{sc}$
are 12 and 200~ms; durations of front edges are 6 and 50~ms at 111
and 61~MHz frequencies, respectively. Additional delay of giant
pulse profile maximum at 61~MHz from 111~MHz is 70~ms (additional
delay of low frequency giant pulses was revealed by \cite{kuz07}
as well, but it was interpreted in other way). According to the
theoretical approach, developed by \cite{wil72} for multiple
scattering, it means that scattering of giant pulses from the Crab
pulsar occurs at thick, and not at thin screen.

\section{Conclusion}

Simultaneous dual-frequency observations of giant pulses from the
Crab pulsar were performed. It is shown that scattering of giant
pulses from the Crab pulsar occurs at thick, and not at thin
screen.

\section{Acknowledgements}

I am grateful to the staff of the Pushchino Radio Astronomy
Observatory for help in preparation and performing of
observations. I am grateful to S.~V. ~Logvinenko for the
development of Pulsar Machine with Fast Fourier Transform. This
paper was submitted as a poster for the 40 Years of Pulsars
Conference. But because of non-arrival of the author the
conference organizers would not let another person to mount this
poster, unfortunately.

%%\clearpage

\begin{figure}
\center
\includegraphics[scale=1.00]{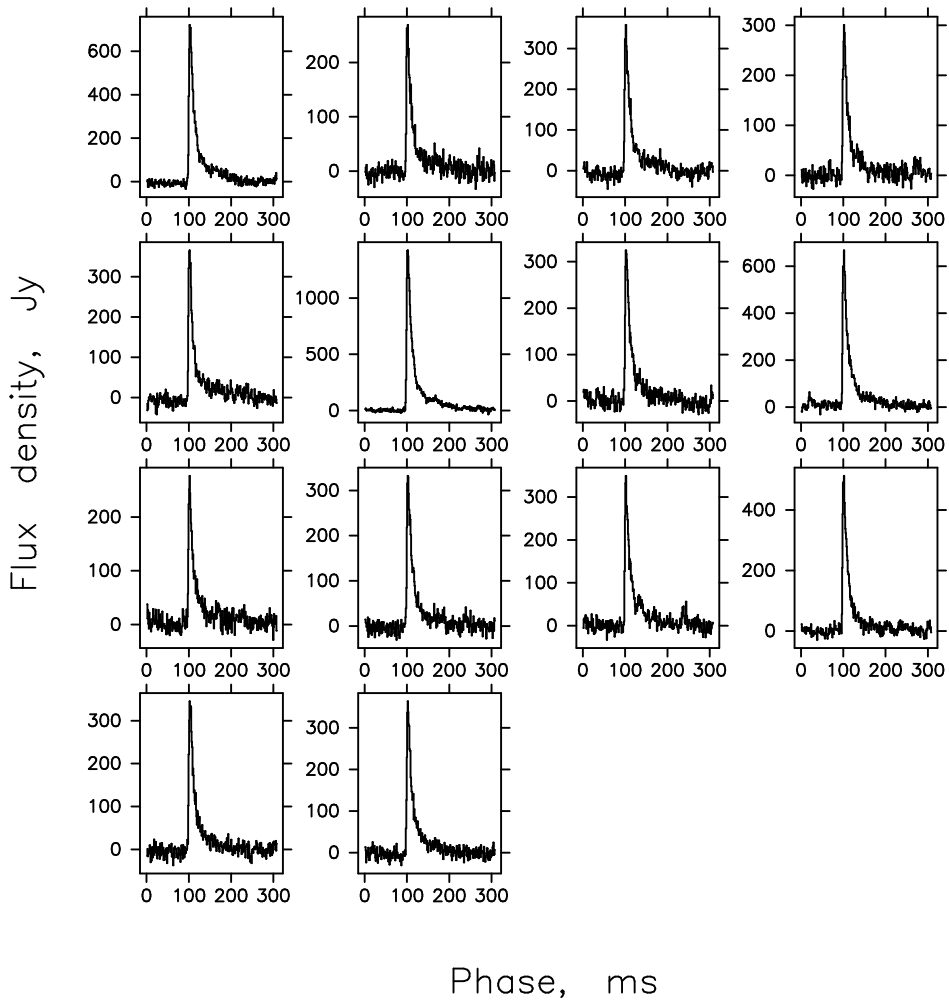}
%\epsscale{.70}
%\plotone{0531_fig1.ps}
\caption{
The profiles of 14 giant pulses from the Crab pulsar at
the frequency of 111~MHz.
\label{fig1}}
\end{figure}

\begin{figure}
\center
\includegraphics[scale=1.00]{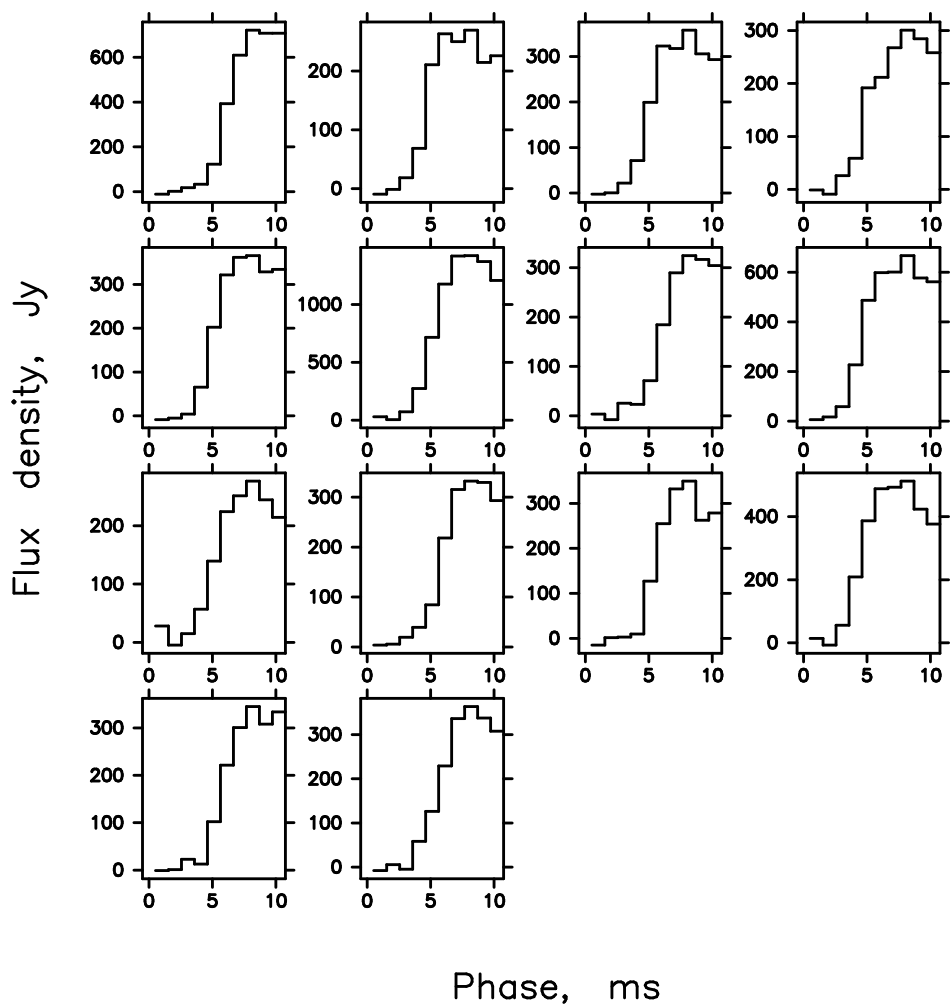}
%\epsscale{.70}
%\plotone{0531_fig2.ps}
\caption{
The rise times of 14 giant pulses from the Crab pulsar
at the frequency of 111~MHz.
\label{fig2}}
\end{figure}

\begin{figure}
\center
\includegraphics[scale=1.00]{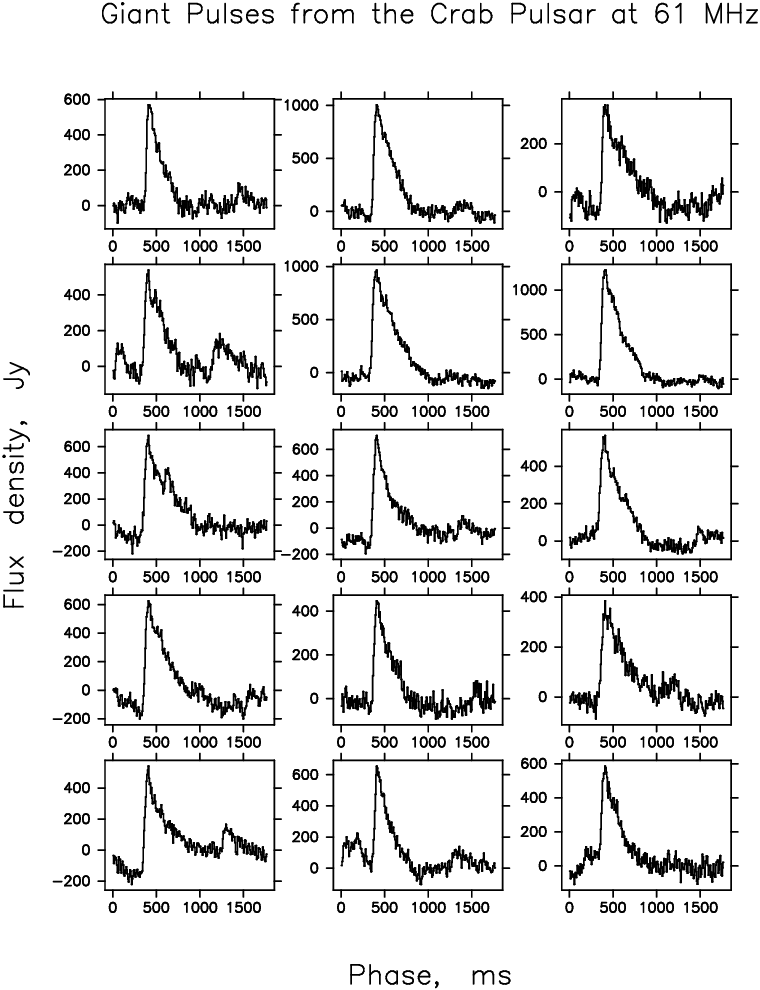}
%\epsscale{.70}
%\plotone{0531_fig3.ps}
\caption{
The profiles of 15 giant pulses from the Crab pulsar at
the frequency of 61~MHz.
\label{fig3}}
\end{figure}

\end{document}